\documentclass[letterpaper,11pt]{article}

\usepackage{scrextend}
\usepackage{amsmath,amssymb,epsfig,bbm}
\usepackage[active]{srcltx}
\usepackage[all]{xypic}
\usepackage{MnSymbol}
\usepackage{slashed}
\usepackage{amsthm}

\usepackage{color}

\usepackage{dsfont}

\definecolor{co}{cmyk}{0,0.7,0.3,0}
\definecolor{darkgreen}{cmyk}{1,0,1,.2}
\definecolor{m}{rgb}{1,0.1,1}

\newcommand{\be}{\begin{equation}}
\newcommand{\ba}{\begin{eqnarray}}
\newcommand{\ea}{\end{eqnarray}}
\newcommand{\nn}{\nonumber}

\def\d{\delta}
\def\e{\epsilon}

\def\m{\mu}
\def\n{\nu}

\def\r{\rho}

\def\ca{{\cal A}}
\def\cb{{\cal B}}

\def\cd{{\cal D}}

\def\cf{{\cal F}}

\def\ch{{\cal H}}

\def\co{{\cal O}}

\makeatletter
\newcommand{\eqnum}{\refstepcounter{equation}\textup{\tagform@{\theequation}}}
\makeatother

\newcommand{\pa}{\partial}

\newcommand{\C}{{\Bbb C}}

\newtheorem{thm}{Theorem}[subsection]

\newtheorem{definition}[thm]{Definition}

\newtheorem*{definition*}{Definition}

\fontfamily{yfrak}

\begin{document}

\vskip 25mm

\begin{center}

{\large\bfseries   On a Lattice-Independent Formulation \\ of Quantum Holonomy Theory
}


\vskip 6ex

Johannes \textsc{Aastrup}\footnote{email: \texttt{aastrup@math.uni-hannover.de}} \&
Jesper M\o ller \textsc{Grimstrup}\footnote{email: \texttt{jesper.grimstrup@gmail.com}}\\ 
\vskip 3ex  


\end{center}

\vskip 3ex

\begin{abstract}

Quantum holonomy theory is a candidate for a non-perturbative theory of quantum gravity coupled to fermions. The theory is based on the $\mathbf{QH D} (M)  $-algebra, which essentially encodes how matter degrees of freedom are moved on a three-dimensional manifold. In this paper we commence the development of a lattice-independent formulation. We first introduce a flow-dependent version of the $\mathbf{QH D}(M)  $-algebra and formulate necessary conditions for a state to exist hereon. We then use the GNS construction to build a kinematical Hilbert space. Finally we find that operators, that correspond to the Dirac and gravitational Hamiltonians in a semi-classical limit, are background independent.

\end{abstract}

\newpage
\tableofcontents

\section{Introduction}
\setcounter{footnote}{0}

Quantum holonomy theory \cite{Aastrup:2015gba} is an attempt to formulate a {\it final} theory, where matter and gauge degrees of freedom are derived from a purely quantum gravitational framework.

%

The theory is based on an elementary algebra, the {\it quantum holonomy-diffeomorphism algebra} denoted $\mathbf{QH D} (M)  $, which is closely related to canonical quantum gravity. This algebra is generated by two types of operators: firstly by local holonomy-diffeomorphisms, which encode how spinors are moved on a 3-dimensional manifold $M$ and which form a non-commutative algebra of functions over a configuration space of connections \cite{Aastrup:2012vq,AGnew}, and secondly by canonical translation operators on this underlying configuration space. Together these two types of operators generate a novel kind of quantum mechanics of diffeomorphisms. The idea is that a fundamental theory can be {\it derived} from this algebra \cite{Aastrup:2005yk}. 

The connection between the $\mathbf{QH D} (M)  $ algebra and canonical quantum gravity is seen in the central algebraic relation of the $\mathbf{QH D} (M)  $ algebra. An infinitesimal version of this relation is identical to the canonical commutation relation of canonical quantum gravity formulated in terms of Ashtekar variables\footnote{In this paper we use $SU(2)$ connections, which in terms of the canonical framework and in terms of Ashtekar connections correspond to either a Euclidian signature or a comparatively more complicated Hamiltonian, see for instance \cite{AL1} for details and further references.}\cite{Ashtekar:1986yd,Ashtekar:1987gu}. This implies that a Hilbert space representation of the $\mathbf{QH D} (M)  $ algebra automatically includes the kinematics of quantum gravity   --  a fact, which in our opinion makes the
 $\mathbf{QH D} (M)  $ algebra an exceedingly natural starting point for a quantum theory of gravity.
 
Moreover, since the holonomy-diffeomorphism part of the $\mathbf{QH D} (M)  $ algebra produces an almost commutative algebra in a semi-classical limit \cite{Aastrup:2015gba} the $\mathbf{QH D} (M)  $ algebra also provides a canonical connection to the research field of non-commutative geometry and in particular to the formulation of the standard model of particle physics coupled to gravity in terms of spectral triples \cite{Connes:1996gi,Chamseddine:2007hz}.

We obtain a kinematical Hilbert space via the GNS construction over a state on a flow-dependent version of the $\mathbf{QH D} (M)  $ algebra, denoted $\mathbf{dQH D}^* (M)  $. The $\mathbf{dQH D}^* (M)  $ algebra combines elements of the $\mathbf{H D} (M)  $ algebra with infinitesimal translation operators from the $\mathbf{QH D} (M)  $ algebra in a certain flow-dependent fashion. The state, that we find, is semi-classical, which means that the kinematical Hilbert space {\it automatically} includes a semi-classical approximation. This feature of quantum holonomy theory is in stark contrast to other non-perturbative approaches to quantum gravity, where the semi-classical approximation remains a significant challenge.

Finally, in \cite{Aastrup:2015gba} we constructed a Hamilton operator, which resembles a curvature operator over the configuration space of Ashtekar connections and from which the Hamilton of general relativity formulated in terms of Ashtekar variables emerges in a semi-classical limit. 

A key crux for a quantum theory of gravity is to check that the constraint algebra is free of anomalies. This determines to what extend the classical symmetries - in this case diffeomorphisms - are preserved in the quantum theory. In \cite{Aastrup:2015gba} we found evidence that the 'Hamilton-Hamilton' sector of the constraint algebra does close off-shell in a non-trivial domain of the Hilbert space. We believe that the computation of the complete constraint algebra is within reach.\\

It thus appears that we are well underway to construct a viable candidate for a theory of quantum gravity -- and indeed, we believe we are. The analysis in \cite{Aastrup:2015gba} was, however, somewhat tarnished by the fact that we used an infinite system of lattice approximations -- the totality of which form a coordinate system -- in order to construct both the $\mathbf{dQH D}^* (M)  $ algebra and the semi-classical state. 

In this paper we commence the development of a lattice-independent formulation of the theory. We first  
formulate the $\mathbf{dQH D}^* (M)  $ algebra and then identify a natural class of states on hereon, which are labeled by a pair of Ashtekar variables and identify necessary conditions for such a state to exist.

In the lattice-independent formulation we find that the $\mathbf{dQH D}^* (M)  $ algebra depends on a flat background metric -- this corresponds to the lattice metric in the lattice-dependent formulation. This metric dependency appears to be an integral part of this framework\footnote{Note, however, that the $\mathbf{QH D} (M)  $ itself does not depend on this background metric.}. We investigate which quantities are independent of this background metric and find that operators, which produce the Dirac Hamiltonian and the gravitational Hamiltonian in the semi-classical limit, do {\it not} depend on this metric.

As was the case in the lattice-dependent formulation we find evidence that the overlap function between different classical geometries vanishes. This suggest that different semi-classical approximations are isolated from each other in terms of quantum transitions, a feature which is a significant departure from standard assumptions about quantum gravity. \\

This paper is organised as follows: In section 2 we introduce both the $\mathbf{QH D} (M)  $ algebra as well as its infinitesimal and flow-dependent version. In section 3 we pause to consider the construction of a Dirac type operator over the configuration space of Ashtekar connections. In \cite{Aastrup:2015gba}  this operator played an important role but we find that the construction of such an operator in a lattice-independent framework is less straight forward. We consider in section 4 the connection to canonical quantum gravity formulated in terms of Ashtekar connections and then move on to construct a state on the $\mathbf{dQH D}^* (M)  $ algebra in section 5. In section 6 we consider the possibility of a state on the $\mathbf{QH D} (M)  $ algebra itself and reproduce the argument first presented in \cite{Aastrup:2015gba} that the overlap function will vanish, which means that no useful state exist. Finally, in section 7 we find that operators, which correspond the classical Hamiltonians, are background independent. Section 8 concludes with a discussion.

\section{Quantum holonomy-diffeomorphism algebras}
\label{firsttask}

We start with the holonomy-diffeomorphism algebra $\mathbf{H D} (M)  $, which was first introduced in \cite{Aastrup:2012vq}, and the quantum holonomy-diffeomorphism algebra $\mathbf{QH D} (M) $  as well as its infinitesimal version $\mathbf{dQH D} (M)  $, which  were introduced in \cite{Aastrup:2014ppa} and \cite{Aastrup:2015gba}.

\subsection{The holonomy-diffeomorphism algebra}
\label{beent}

Let $M$ be a connected $3$-dimensional manifold. Consider the vector bundle $S=M\times \C^2$ over $M$ as well as the space of $SU(2)$ connections acting on the bundle. Given a metric $g$ on $M$ we get the Hilbert space $L^2(M,S,dg)$, where we equip $S$ with the standard inner product. Given a diffeomorphism $\phi:M\to M$ we get a unitary operator $\phi^*$ on  $L^2(M,S,dg)$ via
$$( \phi^* (\xi ))(\phi (m) )= (\Delta \phi )(m)  \xi (m) , $$
where  $\Delta \phi (m)$ is the volume of the volume element in $\phi (m)$ induced by a unit volume element in $ m$ under $\phi $.      

Let $X$ be a vectorfield on $M$, which can be exponentiated, and let $\nabla$ be a $SU(2)$-connection acting on $S$.  Denote by $t\to \exp_t(X)$ the corresponding flow. Given $m\in M$ let $\gamma$ be the curve  
$$\gamma (t)=\exp_{t} (X) (m) $$
running from $m$ to $\exp_1 (X)(m)$. We define the operator 
$$e^X_\nabla :L^2 (M , S, dg) \to L^2 (M ,  S , dg)$$
in the following way:
we consider an element $\xi \in L^2 (M ,  S, dg)$ as a $\C^2$-valued function, and define 
\begin{equation}
  (e^X_\nabla \xi )(\exp_1(X) (m))=  ((\Delta \exp_1) (m))  \hbox{Hol}(\gamma, \nabla) \xi (m)   ,
  \label{chopin1}
 \end{equation}
where $\hbox{Hol}(\gamma, \nabla)$ denotes the holonomy of $\nabla$ along $\gamma$. 
Let $\ca$ be the space of $SU(2)$-connections. We have an operator valued function on $\ca$ defined via 
\begin{equation}
\ca \ni \nabla \to e^X_\nabla  . 
\nn
\end{equation}
We denote this function $e^X$. For a function $f\in C^\infty_c (M)$ we get another operator valued function $fe^X$ on $\ca$. We call this operator a holonomy-diffeomorphisms.

Denote by $\cf (\ca , \cb (L^2(M, S,dg) ))$ the bounded operator valued functions over $\ca$. This forms a $C^*$-algebra with the norm
$$\| \Psi \| =  \sup_{\nabla \in \ca} \{\|  \Psi (\nabla )\| \}, \quad \Psi \in  \cf (\ca , \cb (L^2(M, S,dg )) ). $$

\begin{definition}
Let 
$$C =   \hbox{span} \{ fe^X |f\in C^\infty_c(M), \ X \hbox{ exponentiable vectorfield }\}  . $$
The holonomy-diffeomorphism algebra $\mathbf{H D} (M,S,\ca)   $ is defined to be the $C^*$-subalgebra of  $\cf (\ca , \cb (L^2(M,S,dg )) )$ generated by $C$.

We will often denote $\mathbf{H D} (M,S,\ca)   $ by  $\mathbf{H D}  (M)$ when it is clear which $S$ and $\ca$ is meant.
We will by $\ch \cd (M,S,\ca)   $ denote the  $*$-algebra generated by $C$.
\end{definition}

It was shown in \cite{AGnew} that  $\mathbf{H D} (M,S,\ca)   $ is independent of the metric $g$.

\subsection{The quantum holonomy-diffeomorphism algebra}

Let $\mathfrak{su}(2)$ be the Lie-algebra of $SU(2)$.   
A section $\omega \in \Omega^1(M,\mathfrak{su}(2))$ induces a transformation of $\ca$, and therefore an operator $U_\omega $ on $\mathcal{F}(\ca,  \cb(L^2 (M ,  S,g)))$ via   
$$U_\omega (\xi )(\nabla) = \xi (\nabla - \omega) ,$$ 
which satisfy the relation 
\begin{equation} \label{konj}
(U_{\omega}f e^X U_{ \omega}^{-1}) (\nabla) =f e^X (\nabla + \omega )  , 
\end{equation}
where $f\in C^ \infty_c(M)$. 
Infinitesimal translations on $\ca$ are given by 
\begin{equation}
E_\omega  =\frac{d}{dt}U_{  t  \omega}\Big|_{t=0} \;,
\label{soevnloes}
\end{equation}
where we note that 
$$
E_{\omega_1+\omega_2}=E_{\omega_1}+E_{\omega_2\;,}
$$
which follows since the map $\Omega^1 (M,\mathfrak{su}(2))\ni \omega \to U_{ \omega}$ is a group homomorphism, i.e. $U_{(\omega_1+\omega_2 )}=U_{\omega_1}U_{ \omega_2}$. 

In \cite{Aastrup:2014ppa} we denoted the algebra generated by holonomy-diffeomorphisms and by translations $U_{\omega}$ by $\mathbf{QHD}(M)$ and the algebra generated by holonomy-diffeomorphisms and infinitesimal translations $E_{\omega}$ by $\mathbf{dQHD}(M)$.

Next let $g$ be a metric on $M$ and consider the corresponding isometry $S_g: TM\rightarrow T^*M$. 
With
\begin{equation}
f_y(x) = \left\{
\begin{array}{ll}
1& x=y \\
0& x\not=y
\end{array}\right.
\nn
\end{equation}
we consider the localised operator
$$
E_\omega(x) :=  E_{f_x \omega}\;,
$$
which permit the definiton of the following class of operators. 
Let $\gamma$ be a path in $M$ and $\gamma(t):[a,b]\to M$ be a parametrization of this path. We define
\begin{equation}
E_{\gamma}(t)(\nabla)= Hol ({\gamma_{<t} },\nabla)  \kappa  \sigma^i E_{\sigma^i S_g(X)} (\gamma(t))Hol( {\gamma_{>t} },\nabla)  
\label{omega}
\end{equation}
where $\gamma_{<t}$ be the path $[a,t] \ni \tau\to \gamma(\tau)$ and where $\gamma_{>t}$ be the path $[t,b] \ni \tau\to \gamma(\tau)$. $X$ is a vector field that coincides with $\dot{\gamma}$ on the trajectory of $\gamma$. We define
\begin{equation}
E_{\gamma} (\nabla) =\sum_{t\in [a,b]} E_{\gamma}(t) (\nabla)  ,
\label{FUCK!}
\end{equation}
and check that 
$$
{E}_{\gamma_1} {E}_{\gamma_2} = {E}_{\gamma_1\cdot\gamma_2}\;.  
$$
Equation (\ref{FUCK!}) is essentially the integral of $E_\omega(x)$ along the path $\gamma$, which makes sense since $E_\omega(x) $ transforms as a one-form, except for the important fact that (\ref{FUCK!}) does not have the infinitesimal element '$dt$' and is therefore a formal sum.

An element $F \in\mathbf{HD}(M)$ is a family of operators associated to paths $\{\gamma\}$ in $M$. If we write these operators as $F\vert_\gamma$ then we define the operator $E_F$ is the operator obtained by interchanging each $F\vert_\gamma$ in $F$ with the operator $E_\gamma$. Likewise, we define higher order operators $E_F^{(n)}$ by interchanging operators $F\vert_\gamma$ with operators $E^{(n)}_\gamma$, where the latter is an operators similar to $E_\gamma$ but where $n$ factors of $\kappa  \sigma^i E_{\sigma^i S_g(X)} (\gamma(t))$ are inserted.

 \begin{definition}
We define the $\mathbf{dQHD}^*(M)$ algebra as the $*$-algebra generated by $\mathbf{HD}(M)$ and by all operators $E_F$ and $E_F^{(n)}$. 
\end{definition}
This definition compares to the definition given in the paper \cite{Aastrup:2015gba}, where the $\mathbf{dQHD}^*(M)$ algebra was defined via the interaction between a Dirac type operator and elements in $\mathbf{HD}(M)$, both defined in terms of an infinite system of lattice approximations.

\section{On a Dirac-type operator}

In \cite{Aastrup:2015gba} the construction of a Dirac type operator  {\it over} the configuration space $\ca$ played a key role. It turns out, however, that the construction of such an operator in the lattice-independent formulation is less straight forward. In the following we give a brief outline of the problem.

If we let $(x_1,x_2,x_3)$ denote local coordinates and $g$ be a metric on $M$ then the definition of a Dirac type operator would involve a $\mathfrak{su}(2)$-valued one-form $\mathbbm{e}^i \sigma^i= \mathbbm{e}^i_\m\sigma^i dx^\m$ with an odd grading that satisfies an anti-commutation relation like
\begin{equation}
\left\{   \mathbbm{e}^i_\m(x_1),  \mathbbm{e}^j_\n(x_2) \right\} = \d^{ij}g_{\m\n}(x_1) \d_{x_1x_2},
\label{naborens}
\end{equation}
where $g_{\m\n}$ is the flat metric. 
If we write the operator $\kappa   E_{f_x\sigma_i dx^\m}$ as $\hat{E}_i^\m (x)$ then a plausible definition of a Dirac-type operator would be
$$
D = \sum_{x \in M}  \mathbbm{e}^i_\m(x) \cdot \hat{E}^\m_i(x)\;.
$$
The idea is to obtain relations like
 \begin{eqnarray}
 [D,[D, F]] = 
 E_F, \nn
 \end{eqnarray}
 and 
 \begin{eqnarray}
 [D,\ldots ,[D, F] \ldots ]\Big\vert_{ \mbox{\tiny $2n$ commutators}} = 
 E^{(n)}_F,
 \label{vesterhau}
 \end{eqnarray}
 with $F\in \mathbf{HD}(M)$. 
 The problem, that one encounters, however, is that these relations entails operators $E_F$  and $E^{(n)}_F$ that involve {\it line integrals} instead of the sums as defined in (\ref{FUCK!}). This, in turn, completely changes the algebraic structure of the 
$\mathbf{dQHD}^*(M)$ algebra. The problems boils down to finding the right defintion of the Clifford algebra in (\ref{naborens}). Since we are at the moment uncertain as to what definition of a Dirac type operator is most suitable we shall here simply leave this question unresolved and make do with the definition of the $E_F$  and $E^{(n)}_F$ operators given in the previous section.

The reason why we seek to define a Dirac type operator over the space $\ca$ is that this would represent a natural geometrical structure. We believe that such a structure - be it a Dirac type operator or something similar - is desirable in order to have a guiding principle in the definition of the theory. 


\section{Canonical quantum gravity}

Before we continue the analysis of the $\mathbf{QHD}(M)$ and $\mathbf{dQHD}^*(M)$ algebras let us for a moment pause to consider their connection to canonical quantum gravity.

If we combine equation (\ref{konj}) with (\ref{soevnloes}) we obtain
\begin{equation} \label{flowkan}
[ E_\omega , e^X_\nabla ]= \left.\frac{d}{dt}e^X_{\nabla +t\omega}\right\vert_{t=0}  . 
\end{equation}
To analyse the righthand side of (\ref{flowkan}) we introduce local coordinates $(x_1 ,x_2,x_3)$. We decompose $\omega$: $\omega =\omega^i_\m\sigma_i dx^\m$. 
For a given point $p\in M$ choose the points $$p_0=p,\quad p_1=e^{\frac{1}{n}X}(p),\ldots  ,\quad  p_n= e^{\frac{n}{n}X}(p)$$ 
on the path
$$t\to e^{tX}(p)  ,t\in [0,1].$$
We write the vectorfield $X=X^\n\partial_\n$. 
We have 
\begin{eqnarray} 
\lefteqn{e^X_{\nabla+t\omega}}\nn\\
& =&\lim_{n\to \infty } (1+\frac{1}{n}(A(X(p_0))+t \omega^i_\m \sigma_i X^\m(p_0) ) (1+\frac{1}{n}(A(X(p_1))+t \omega^i_\m \sigma_iX^\m(p_1))\nn\\
&& \cdots (1+\frac{1}{n}(A(X(p_n))+t \omega^i_\m \sigma_i X^\m(p_n)) ,
\label{vlad}
\end{eqnarray}
where  $\nabla=d+A$, and therefore 
\begin{eqnarray}
 \lefteqn{\frac{d}{dt}e^X_{\nabla +t\omega}\Big\vert_{t=0}}
 \nn
 \\
 &=& \lim_{n\to \infty }  \Big( \frac{1}{n} \omega^i_\m\sigma_iX^\m(p_0)  (1+\frac{1}{n}A(X(p_1)))\cdots (1+\frac{1}{n}A(X(p_n))) \nn\\
 &&+ (1+\frac{1}{n}A(X(p_0))) \frac{1}{n}  \omega^i_\m \sigma_iX^\m(p_1)  (1+\frac{1}{n}A(X(p_2)))\cdots (1+\frac{1}{n}A(X(p_n))) \nn\\
 && + \quad\quad\quad\quad\quad\quad\quad\quad\quad\quad\quad \vdots \nn\\
 &&+ (1+\frac{1}{n}A(X(p_0)))   (1+\frac{1}{n}A(X(p_2)))\cdots 
 \nn\\
 &&\hspace{4cm}\cdots
 (1+\frac{1}{n}A(X(p_{n-1}))) \frac{1}{n}  \omega^i_\m \sigma_i X^\m(p_n) \Big)
\label{COOM}
\end{eqnarray}
This equation should be compared to the classical setup of Ashtekar variables and holonomies of Ashtekar connections \cite{AL1}. There we have canonically conjugate variables $(\mathds{E}^\m_i , \mathds{A}_\n^j)$ where indices $\{i,j,k,...\}$ are $\mathfrak{su}(2)$ indices and $\{\m,\n,...\}$ are indices labelling a coordinate system on $M$. $\mathds{E}$ is a densitized inverse triad field $\mathds{E}^\m_i= e e^\m_i$ where $e^\m_i$ is the inverse triad field and $e$ its determinant. $\mathds{A}$ is the Ashtekar connection\footnote{note that we here work with $SU(2)$ connections which correspond to either a Euclidian signature or a Hamiltonian with a comparably more complicated structure, see for instance \cite{AL1}. }. If one considers instead of $\mathds{E}$ its flux over a two-surface $S$
\begin{equation}
F^S_i =\int_S \e_{\m\n\r}  \mathds{E}_i^\m dx^\n dx^\r
\label{flux}
\end{equation}
then the Poisson bracket between the holonomy of $\mathds{A}$ along a curve $\gamma$ and $F^S_i$ reads \cite{AL1}
$$
\left\{  F^S_i, Hol(\gamma,\mathds{A})   \right\}_{PB} = \iota(S,\gamma)  Hol(\gamma_1,\mathds{A}) \sigma_i Hol(\gamma_2,\mathds{A})  
$$
where $\gamma = \gamma_1 \cdot \gamma_2$ and where the Pauli matrix is inserted at the point of intersection between $S$ and $\gamma$. $\iota(S,\gamma)=\pm 1$ or $0$ encodes information on the intersection of $S$ and $\gamma$.   

\begin{figure}[t]
\begin{center}
\resizebox{!}{ 4 cm}{
 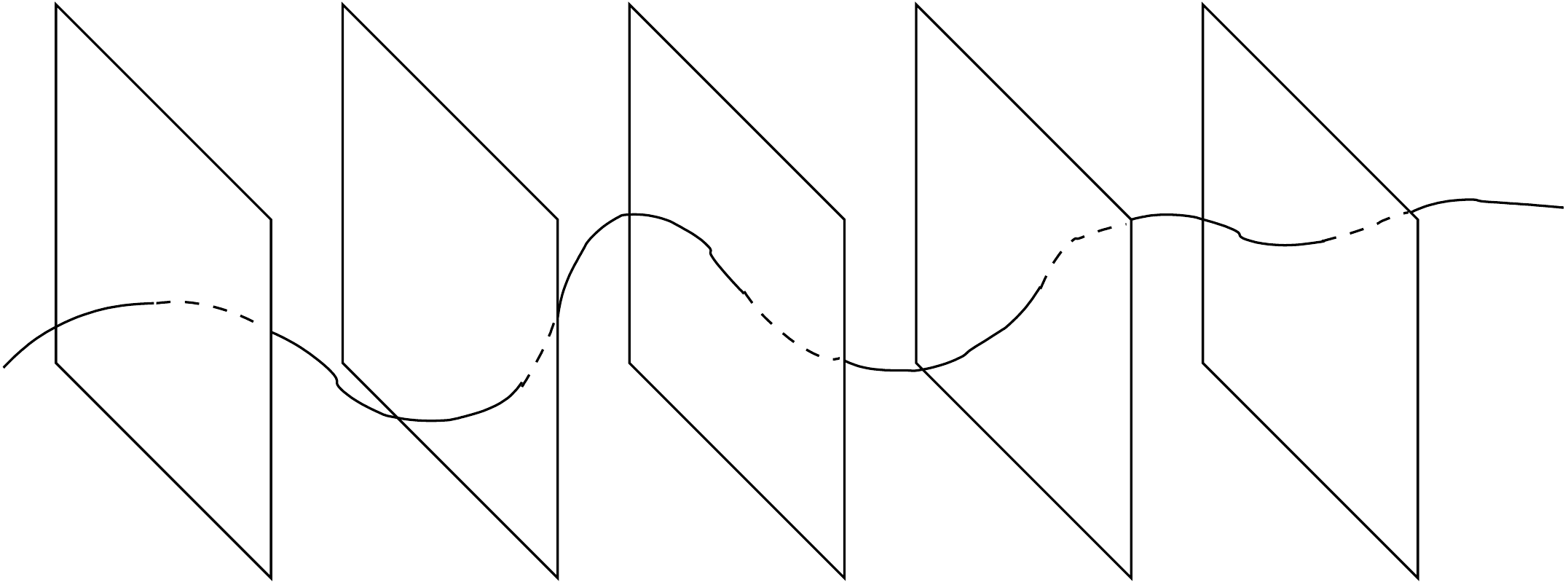}
\end{center}
\caption{\it The operator $E_{\sigma_i dx^\m}$ will, when it is commuted with a flow, insert Pauli matrices continuously along the course of the flow. This means that it acts as a sum of flux operators with surfaces, which intersect the flow at the points of insertion. }
 \label{heltvildtsvedig}
\end{figure}

We therefore see that before taking the limit $\lim_{n\to \infty}$ of (\ref{COOM}) we have simply the commutator of the sum of the flux operators $\sum_k \frac{1}{n}X^\m (p_k) F^{S_k}_i $, where $S_k$ is the plane orthogonal to the $x_\m$-axis intersecting $p_k$, and the holonomy operator of the path $$t\to e^{tX}(p)  ,t\in [0,1],$$
 see figure \ref{heltvildtsvedig}.
 
It follows that $E_{\sigma_i dx^\m}$ is  a series of flux-operators $F^S_i$ sitting along the path $$t\to e^{tX}(p)  ,t\in [0,1],$$
where the surfaces $S$ are just the planes othogonal to the $x_\m$ direction.
But since there are infinitely many of them, they have been weighted with the infinitesimal length, i.e. with a $dx^\m$, see figure \ref{heltvildtsvedig}. We can formally write this as
$$
E_{\sigma_i dx^\m} = \int \hat{F}^S_i dx^\m  \;, 
$$
where $ \hat{F}^S_i$ is an operator, which corresponds to a quantization of the flux operator (\ref{flux}).   
This provides us with a solid interpretation of the $\mathbf{QHD}(M)$ algebra in terms of canonical quantum gravity, where the operator $E_\omega$ is a global flux operator.

Finally note that the phenomenon from the holonomy-flux-algebra, that a path $p$ running inside a surface $S$, has zero commutator with the corresponding flux operator is encoded in the $\mathbf{QHD}(M)$ algebra, since the tangent vectors of $p$ will be annihilated by the differential form $dx^\m$.

We can also make the holonomies infinitesimal in order to see the canonical commutation relations between the Ashtekar variables directly. This was done in \cite{Aastrup:2015gba} and shall not be done here.

\section{States on $\mathbf{HD}(M)$ and $\mathbf{dQHD}^*(M)$}

In the following we shall think of states on algebras that involve $\mathbf{HD}(M)$ or $\mathbf{dQHD}^*(M)$ in terms of maps
\begin{equation}
\rho:\mbox{algebra}\rightarrow M_2(\mathbb{C})\otimes \cf(M\times M)
\label{morgensang}
\end{equation}
with the understanding that we obtain a state by composing $\rho$ with one of these maps
\begin{eqnarray}
&&\Phi_{\psi}: M_2(\mathbb{C})\otimes \cf(M\times M) \rightarrow \mathbb{C}\; , 
\nn\\
&&\hspace{0.8cm} K(x,y)   \rightarrow   \int_{M\times M}   Tr_{M_2}  (\bar{\psi}(x) K(x,y)  \psi(y))dxdy
\nn\\
&&\Phi_{\mbox{\tiny vac}}: M_2(\mathbb{C})\otimes \cf(M\times M) \rightarrow \mathbb{C}\; , 
\nn\\
&&\hspace{0.8cm} K(x,y)   \rightarrow   \int_{M\times M}   Tr_{M_2} (  K(x,y))  \d^{3}(x-y) dxdy
\label{thy}
\end{eqnarray}
where $\psi$ is a $\mathbb{C}^2$-valued function on $M$, i.e. a spinor. Note that these maps depend on a measure on $M$. 

We are now ready to write down the state on $\mathbf{dQHD}^*(M)$. Let ${\bf A}$ be a map from $TM$ to $M_2(\C)$, whose properties will be specified shortly, and ${\bf E}= {\bf E}^\m_i \sigma^i \pa_\m$ be an element in $TM$ that takes values in $\mathfrak{su}(2)$. 
We specify the map $\rho_{({\bf A},{\bf E})}^\kappa$ first with 
\begin{eqnarray}
\rho_{({\bf A},{\bf E})}^\kappa( fe^X) (x,y) = f(x) Hol(\gamma,{\bf A}) \d^{3}(y-\exp(X)(x)) 
\label{hvisker1}
\end{eqnarray}
where $\gamma$ is the curve in $M$ generated by $X$. The function $f$ is evaluated at the start-point of $\gamma$. 

Recall that $E_F$ restricted to a path $\gamma$ is denoted by $E_\gamma$ as defined in (\ref{omega}). Thus, we need to specify the map $\rho_{({\bf A},{\bf E})}^\kappa$ on operators, which, when restricted to a path $\gamma=\gamma_1\cdot\gamma_2\cdot\ldots\gamma_{n+1}$, have the general structure
\begin{eqnarray}
 Hol(\gamma_1,\cdot) \hat{E}^{i_1} (x_1) \sigma^{i_1} Hol(\gamma_2,\cdot)    \hat{E}^{i_2} \sigma^{i_2} (x_2) \ldots    \hat{E}^{i_n} \sigma^{i_n} (x_n)  Hol(\gamma_{n+1},\cdot)  
\end{eqnarray}
where each operator of $ {E}^i (x_1) \sigma^i $ is inserted along a section of the path $\gamma$. The map $\rho_{({\bf A},{\bf E})}^\kappa$ is then defined via the left-ordering
\begin{eqnarray}
 &&\hspace{-10mm}\rho_{({\bf A},{\bf E})}^\kappa\left(  Hol(\gamma_1,\cdot)  \hat{E}^{i_1} (x_1) \sigma^{i_1} Hol(\gamma_2,\cdot)    \hat{E}^{i_2} \sigma^{i_2} (x_2) \ldots    \hat{E}^{i_n} \sigma^{i_n} (x_n)  Hol(\gamma_{n+1},\cdot)    \right)
\nn\\
&=& \rho_{({\bf A},{\bf E})}^\kappa\left(    \hat{E}^{i_1} (x_1)  \hat{E}^{i_2}  (x_2)  \ldots   \hat{E}^{i_n} (x_n)  \right.
\nn\\
&&\hspace{2cm}\cdot\left.
Hol(\gamma_1,\cdot)  \sigma^{i_1} Hol(\gamma_2,\cdot)   \sigma^{i_2}\ldots   \sigma^{i_n}   Hol(\gamma_{n+1},\cdot)   
 \right)
 \nn\\
 &&+ \mbox{lower order terms}
\end{eqnarray}
where 'lower order terms' refer to terms, which arise from commuting operators $E_\omega(x)$ through holonomy-diffeomorphisms. Thus, these lower order terms all involve pairs of contracted Pauli matrices and are of order $\kappa$ or higher. We continue with
\begin{eqnarray}
&&\hspace{-10mm} \rho_{({\bf A},{\bf E})}^\kappa\left(    \hat{E}^{i_1} (\gamma_{j_1}(t_{j_1}))  \hat{E}^{i_2}  (\gamma_{j_2}(t_{j_2})) \ldots   \hat{E}^{i_n}  (\gamma_{j_n}(t_{j_n})) \right.
\nn\\
&&\hspace{2cm}\cdot\left.
Hol(\gamma_1,\cdot)  \sigma^{a_1} Hol(\gamma_2,\cdot)   \sigma^{a_2}\ldots   \sigma^{a_m}   Hol(\gamma_{m+1},\cdot)   
 \right)
 \nn\\
 &=&
    {\bf E}^{i_1}  (\gamma_{j_1}(t_{j_1})) dt _{j_1} {\bf E}^{i_2}  (\gamma_{j_2}(t_{j_2}))dt _{j_2}  \ldots  {\bf E}^{i_n}  (\gamma_{j_n}(t_{j_n}))dt _{j_n} 
\nn\\
&&\hspace{2cm}\cdot
Hol(\gamma_1, {\bf A} )  \sigma^{a_1} Hol(\gamma_2,{\bf A})   \sigma^{a_2}\ldots   \sigma^{a_m}   Hol(\gamma_{m+1},{\bf A})   
\end{eqnarray}
Note that the state provides each '$ {\bf E}^{i_k}$' with an infinitesimal element '$dt$', which ensures that the formal sum in (\ref{FUCK!}) is converted into a line integral.

Next, we write ${\bf A}$ as 
$$
{\bf A} = \mathds{A} + \mathds{A}_q
$$
where $\mathds{A}$ is a one-form, which takes values in $\mathfrak{su}(2)$ and where 
$$ 
\mathds{A}_q= \sum_{n=1}^\infty \kappa^n \mathds{A}^{(n)}
$$ 
is a map $TM\rightarrow M_2(\mathbb{C})$ that satisfies the following homogeneity condition
$$
\mathds{A}_q(\lambda X) = \vert \lambda \vert \mathds{A}_q( X)\;,\quad \lambda\in  \mathbb{C}
$$
and where $\mathds{A}_q( X) $ is a negative definite element in $M_2(\mathbb{C})$.
Here $\kappa$ enters as a quantization parameter, which separates the classical contribution from its 'quantum' counterparts. Accordingly, we may expand ${\bf E}$ in
$$
{\bf E} = \mathds{E} + \mathds{E}_q\;,
$$
with $\mathds{E}_q= \sum_{n=1}^\infty   \kappa^{n} \mathds{E}^{(n)}$.
At this point ${\bf E}$ is not restricted but we will shortly see that both ${\bf A}$ and ${\bf E}$ are required to meet an infinite sequence of requirements in order for $ \xi_{({\bf A},{\bf E})}^\kappa$ to give rise to a state. Before we do that we note that $(\mathds{A},\mathds{E})$ form a set of Ashtekar variables\footnote{note again that we have $SU(2)$ connections which correspond to either a Euclidian signature or a Hamiltonian with a more complicated structure, see \cite{AL1}. } \cite{Ashtekar:1986yd,Ashtekar:1987gu}, i.e. a point in the phase space of canonical quantum gravity.

The complete state on $\mathbf{dQHD}^*(M)$ is then written in one of two ways:
\begin{eqnarray}
\rho^\kappa_{(\psi,{\bf A},{\bf E})}&=&\Phi_{\psi}\circ \rho_{({\bf A},{\bf E})}^\kappa\;,\quad
\nn\\
\rho^\kappa_{\mbox{\tiny vac}({\bf A},{\bf E})}&=&\Phi_{\mbox{\tiny vac}}\circ \rho_{({\bf A},{\bf E})}^\kappa\;.
\label{tristese}
\end{eqnarray}
In \cite{Aastrup:2015gba} we find that these maps give rise to a state on $\mathbf{HD}(M)$. The analysis there was carried out in terms of an infinite system of lattice approximations. See also \cite{jubi} where we present a more thorough analysis.  We are now going to check whether we also have states on $\mathbf{dQHD}^*(M)$. We start with $\rho^\kappa_{\mbox{\tiny vac}({\bf A},{\bf E})} $ and write
\begin{equation}
\rho^\kappa_{\mbox{\tiny vac}({\bf A},{\bf E})} (a a^*) = {Z} + {W}
\label{sanders}
\end{equation}
where $a= \sum_i c_i a_i \in\mathbf{dQHD}^*(M)$ is an arbitrary element in the algebra where each $a_i$ is an operator $fe^X$ or $F_{E}^{(n)}$, where $c_i$ are constants and where ${Z} $ is the positive real number one would obtain if $ \hat{E}^\m_i $ had commuted with elements in $\mathbf{HD}(M)$. $W$ is then the term generated by commutators between operators $ \hat{E}^\m_i $ and elements in $\mathbf{HD}(M)$. We need to show that $W$ is real and that $\vert W\vert < Z$. 

We start with $a= E_F$ and consider a single path $\gamma$ in $F$, which is parametrised so that $\gamma(t_s)$ and $\gamma(t_e)$ is the start and end-point of the path. For simplicity we assume that $F$ does not self-intersect and we write $Z$ and $W$ as two-by-two matrices tensored with functions on $M\times M$ and thereby avoiding a trace and the spinor. In this case we have:
\begin{eqnarray}
Z &=&
\left( \int dt  Hol ({\gamma_{<t} },{\bf A})   g_{\m\n}\sigma^i {\bf E}_i^{\m} dx^\n (\gamma(t))   Hol( {\gamma_{>t} },{\bf A}) \right) \nn\\
&&\cdot
   \left( \int dt  Hol ({\gamma_{<t} },{\bf A})   g_{\m\n}\sigma^i {\bf E}_i^{\m}  dx^\n (\gamma(t))   Hol( {\gamma_{>t} },{\bf A}) \right)^*\;,
 \nn\\
 W&=& 
 \kappa
\int_0^1 dt' \int_{0}^{t'} dt  Hol ({\gamma_{<t} },{\bf A})   g_{\m\n}\sigma^i {\bf E}_i^{\m} dx^\n (\gamma(t))  Hol ({\gamma_{t<t'} },{\bf A})   \nn\\
&&\cdot
  \vert \dot{\gamma} \vert^2(t')   \sigma^j  Hol( {\gamma_{>t'} },{\bf A})    Hol^* ({\gamma_{>t'} },{\bf A})   \sigma^j Hol^* ( {\gamma_{<t'} },{\bf A}) 
 \nn\\
 &&+      \kappa 
 \int_0^1 dt' \int^{1}_{t'} dt   Hol ({\gamma_{<t'} },{\bf A})   \vert \dot{\gamma} \vert^2(t')    \sigma^j     Hol ({\gamma_{t'<t} },{\bf A})        \nn\\
&&\cdot
     g_{\m\n}\sigma^i {\bf E}_i^{\m} dx^\n (\gamma(t))  Hol( {\gamma_{>t} },{\bf A})   Hol^* ({\gamma_{>t'} },{\bf A})   \sigma^j Hol^* ( {\gamma_{<t'} },{\bf A}) 
\label{sommertime}
\end{eqnarray}
where $\gamma_{t<t'}$ be the path $[t,t'] \ni \tau\to \gamma(\tau)$ and where $\vert \cdot \vert $ is with respect to the metric $g$.
In order to ease the notation we are going to proceed with the shorthand notation:
$$
Z= \left\vert \stackrel{\bf A}{\Longleftrightarrow } {\bf E}\stackrel{\bf A}{\Longleftrightarrow }\right\vert^2 \;,\quad W =  \left(\stackrel{\bf A}{\Longleftrightarrow } {\bf E}\stackrel{\bf A}{\Longleftrightarrow }\right) \left(\sigma \stackrel{\bf A}{\Longleftrightarrow }\sigma \stackrel{\bf A}{\Longleftrightarrow }\right)^*\;.
$$
We are going to consider the requirement $Z>\vert W\vert$ at increasing orders in $\kappa$ starting with the classical level and going up to second order. In this way we will find sufficient but not exhaustive conditions for a state to exist.\\

\underline{Zeroth order in $\kappa $:}

At this level $W$ does not contribute and thus the condition $Z>\vert W\vert$ is trivially met. This means that this analysis has no implications for the classical setup.\\

\underline{First order in $\kappa $:}
\begin{eqnarray}
Z &=& \left( \stackrel{ \mathds{A}^{(1)}}{\Longleftrightarrow } { \mathds{E} }\stackrel{ \mathds{A}}{\Longleftrightarrow }\right)\left( \stackrel{ \mathds{A}}{\Longleftrightarrow } {  \mathds{E} }\stackrel{ \mathds{A}}{\Longleftrightarrow }\right)^* + \left( \stackrel{ \mathds{A}}{\Longleftrightarrow } {  \mathds{E} ^{(1)}}\stackrel{ \mathds{A}}{\Longleftrightarrow }\right)\left( \stackrel{ \mathds{A}}{\Longleftrightarrow } {  \mathds{E} }\stackrel{ \mathds{A}}{\Longleftrightarrow }\right)^*
\nn\\&&
+\left( \stackrel{ \mathds{A}}{\Longleftrightarrow } {  \mathds{E} }\stackrel{ \mathds{A}^{(1)}}{\Longleftrightarrow }\right)\left( \stackrel{ \mathds{A}}{\Longleftrightarrow } {  \mathds{E} }\stackrel{ \mathds{A}}{\Longleftrightarrow }\right)^* + \ldots
\nn\\
W &=&  \left(\stackrel{ \mathds{A} }{\Longleftrightarrow } { \mathds{E} }\stackrel{\mathds{A}}{\Longleftrightarrow }\right) \left( \sigma\stackrel{\mathds{A}}{\Longleftrightarrow } \sigma \stackrel{\mathds{A}}{\Longleftrightarrow }\right)^* 
\end{eqnarray}
where $Z$ has six terms. At this level in $\kappa$ the requirement $Z> \vert W\vert$ puts restrictions on $\mathds{E}^{(1)}$ and $\mathds{A}^{(1)} $ in terms of the classical fields $\mathds{E}$ and $\mathds{A}$.\\

\underline{Second order in $\kappa $:}
\begin{eqnarray}
Z &=& \left( \stackrel{ \mathds{A}^{(2)}}{\Longleftrightarrow } {  \mathds{E} }\stackrel{ \mathds{A}}{\Longleftrightarrow }\right)\left( \stackrel{ \mathds{A}}{\Longleftrightarrow } {  \mathds{E} }\stackrel{ \mathds{A}}{\Longleftrightarrow }\right)^* + 
\left( \stackrel{ \mathds{A}}{\Longleftrightarrow } {  \mathds{E} ^{(2)}}\stackrel{ \mathds{A}}{\Longleftrightarrow }\right)\left( \stackrel{ \mathds{A}}{\Longleftrightarrow } {  \mathds{E} }\stackrel{ \mathds{A}}{\Longleftrightarrow }\right)^* + \ldots
\nn\\&&
+ \left( \stackrel{ \mathds{A}^{(1)}}{\Longleftrightarrow } {  \mathds{E}^{(1)}}\stackrel{ \mathds{A}}{\Longleftrightarrow }\right)\left( \stackrel{ \mathds{A}}{\Longleftrightarrow } {  \mathds{E} }\stackrel{ \mathds{A}}{\Longleftrightarrow }\right)^* + \ldots
\nn\\
W &=&  \left(\stackrel{ \mathds{A}^{(1)} }{\Longleftrightarrow } { \mathds{E} }\stackrel{\mathds{A}}{\Longleftrightarrow }\right) \left( \sigma \stackrel{\mathds{A}}{\Longleftrightarrow } \sigma \stackrel{\mathds{A}}{\Longleftrightarrow }\right)^* +  \left(\stackrel{ \mathds{A}}{\Longleftrightarrow } { \mathds{E}^{(1)} }\stackrel{\mathds{A}}{\Longleftrightarrow }\right) \left( \sigma \stackrel{\mathds{A}}{\Longleftrightarrow } \sigma \stackrel{\mathds{A}}{\Longleftrightarrow }\right)^* \nn\\&& + \ldots
\end{eqnarray}
where $Z$ has twelve terms and $W$ has five terms. Again, at this level in $\kappa$ the requirement $Z> \vert W\vert$ puts restrictions on $\mathds{E}^{(2)}$ and $\mathds{A}^{(2)} $ in terms of $\mathds{E}^{(1)}$ and $\mathds{A}^{(1)} $ as well as the classical fields $\mathds{E}$ and $\mathds{A}$.

In the case where $a$ in (\ref{sanders}) is a sum over different $E_F$'s the $W$ in (\ref{sommertime}) will have a slightly different form in the sense that: 1) for those terms that involve different $F$'s the contribution will only be non-zero if they intersect and 2) the insertion of $\sigma$-matrices only happens in those sections of the holonomy-diffeomorphisms where the different $E_F$'s intersect. Otherwise the structure of $W$ is the same.

In the case where we consider (sums of) higher order operators $E_F^{(n)}$ the picture becomes more complicated but nevertheless still has the same overall structure: terms in $W$ in (\ref{sommertime}) will involve terms where $\hat{E}$'s are commuted through sections of holonomy-diffeomorphisms where $\sigma$-matrices are then inserted.

It is beyond the scope of this paper to fully analyse these conditions. It seems plausible that an 'order-by-order' approach is feasible, where at the $n$'th order conditions for the terms $\mathds{A}^{(n)}$ and $\mathds{E}^{(n)}$ are determined by lower-order terms. The question is then whether the sums of these terms converge. 
We do, however, not see any reason why these conditions should not have solutions. One argument in favour of solutions is the following: the above computation is essentially one-dimensional and therefore also apply for a construction where holonomies are restricted to various forms of lattices. One example of such a construction is the formulation of QHT used in \cite{Aastrup:2015gba} and another example is Loop Quantum Gravity \cite{AL1}. In the latter case states are known to exist (see for example \cite{BT2} and references therein).  

For the remainder of this paper we shall assume that these condition can be met and that the maps in (\ref{tristese}) are states for appropriate choices of ${\bf A}$ and ${\bf E}$.

With the GNS construction over the state constructed in the above the Dirac type operator will not be an operator in the ensuing Hilbert space nor does it seem possible to enlarge to Hilbert space to encompass this operator. This appears to be a generic characteristics, which we also encountered in \cite{Aastrup:2015gba}, where our analysis was based on lattice approximations. The reason is this: the state, which we find, assigns each operator $\hat{E}$ with an infinitesimal line element $dt$, which is necessary in order for the interaction between $D$ and elements in $\mathbf{HD(M)}$ to be well defined. On the other hand, if $D$ itself -- and not merely its commutators with elements in $\mathbf{HD(M)}$ -- should exist as an operator in the GNS construction this assignment would have to be different (see  \cite{Aastrup:2015gba} for details and discussion).

\section{Some properties of states on $\mathbf{QHD}(M)$}

At this point in our analysis one may ask why we do not work with states on the $\mathbf{QHD}(M)$-algebra instead of states on $\mathbf{dQHD}^*(M)$. After all, the former appear to be the canonical choice of an algebra. The answer, which we discussed also in \cite{Aastrup:2015gba} and \cite{Aastrup:2014ppa}, is that there does not seem to be any reasonable states on $\mathbf{QHD}(M)$. The reason for this is the following: A state of the form
$$
\rho_{\bf A} \left(fe^X\right)(x,y)=Hol (\gamma , {\bf A}) f(s(\gamma )) \delta^3(y-\exp (X )(x))\;,
$$
has the property that  $Hol (\gamma , {\bf A})$ is a matrix of norm $<1$ and $>0$ and the norm is decreasing the longer $\gamma$ gets. If we therefore look at $U_{\omega}$ restricted to $\gamma$, we must have $|\rho_{\bf A} (U_\omega|_\gamma)|<1$ when $\omega \not= 0$ on $\gamma$, because if $|\rho_{\bf A} (U_\omega|_\gamma)|=1$, the state will be equally distributed on the space of connections $\ca\vert_\gamma$, and hence one would have $Hol (\gamma , {\bf A})=  0$. On the other hand if  $|\rho_{\bf A} (U_\omega|_\gamma)|<1$ on each $\gamma$, we will have $|\rho_{\bf A} (U_\omega)|=0$, since we have 
$$
|\rho_{\bf A}(U_\omega )|=|\rho_{\bf A}(U_\omega|_\gamma)||\rho_{\bf A}(U_\omega|_{M\setminus \gamma})|\;.
$$ 
We can continue to pull out another path of $M\setminus \gamma$, and get another factor $<1$. In this way $|\rho_{\bf A}(U_\omega )|$ becomes an infinite product of factors $<1$, and hence $\vert \rho_{\bf A}(U_\omega )\vert=0$.

The validity of this argument depends on a number of assumptions. First of all there is the assumption that we can restrict the $U_\omega$ to subsets, especially one-dimensional subsets. Another assumption is, that with respect to this restriction the expectation values behaves functorial, i.e.    $|\rho_{\bf A}(U_\omega )|=|\rho_{\bf A}(U_\omega|_\gamma)||\rho_{\bf A}(U_\omega|_{M\setminus \gamma})|$. 

There are of course states, which do not possess this property. For instance we could take the state coming from the direct sum of two representations of $\mathbf{QHD}(M)$ given by two different connections $\nabla_1$ and $\nabla_2$. In this case the state on $U_{\nabla_1-\nabla_2}$ will not be zero. However this type of state is not of the form $ \rho_{\bf A}$.

A different type of state, which is also not covered by the above argument is the state $\rho_{\bf \infty}$ given by 
\begin{equation}
\rho_{\bf \infty} \left(fe^X\right)(x,y)=0\; \hbox{ for }X(x)\not= 0,\quad \rho_{\bf \infty} (U_\omega)=1 .
\label{highenergy}
\end{equation}
This is a type of state, which is equally distributed on all connections. 

A variant on this type of state, is a state which looks like $\rho_{\bf \infty}$  apart from on a single path, where it looks like $\rho_{\bf A}$ for some $\bf A$. These types of states are also not covered by the above argument. It is unclear whether these give states on the $\mathbf{HD}(M)$ algebra, or just the flow-algebra (see \cite{Aastrup:2015gba} for details on the flow-algebra), since a path has measure zero on $M$ when the dimension of $M$ is bigger than one. 

On might speculate whether the state in (\ref{highenergy}) represents a high-energy state, where all notions of geometry have vanished in the sense that the state is equally distributed on the space $\ca$, and whether there could be a kind of phase transition from this state into the type of states described in the previous section, which are localised around a single classical geometry. 

It seems to be a feature of all these states described here, that they are in one way or another not strongly continuous. The states $\rho_{\bf A}$ are not strongly continuous on the $\mathbf{dQHD}^*(M)$-algebra due the the singular nature of the commutator in $\mathbf{dQHD}^*(M)$. The other types of states are not strongly continuous on $\mathbf{HD}(M)$. The lack of strongly continuity makes it difficult to consider infinitesimal objects.

If the above argument holds then it has surprising consequences for this approach to a theory of quantum gravity. It means that the overlap function between different classical geometries is strictly zero. Thus, each semi-classical approximations is isolated from other semi-classical approximation. This does, however, {\it not} imply that this approach is essentially classical. As we discussed in section 7 in \cite{Aastrup:2015gba} there exist states in the GNS construction discussed above, which are highly non-classical.

\section{The Hamiltonians}

The algebra $\mathbf{dQHD}^*(M)$ depends on the background metric $g$. In the following we let $g_{\m\n}=\d_{\m\n}$ be the flat metric. We will se that important operators such as those, which give rise to the Dirac Hamiltonian in the semi-classical limit, do not depend on this background structure.

We will first analyse the classical limit of an operator, which we shall see corresponds to a quantization of a spatial Dirac operator. We set $\kappa=0$ and 
 introduce local coordinates $(x_1,x_2,x_3)$. We write down the expectation value of the derivative of a holonomy acting on a spinor $\psi$
$$
\frac{d}{ds}\left( \xi^\kappa_{({\bf A},{\bf E})}\left(e^{sX}\right)  \psi \right) \Big\vert_{s=0,\kappa=0}  (x)  =  \left( X(x)+\mathds{ A}(X(x))\right)\psi(x) \;.
$$
Next we consider the following operator, that involves both $\sigma^iE_{\sigma^i S_g(X)}(x)$ and a holonomy-diffeomorphism
\begin{eqnarray}
\frac{d}{ds}\left( \xi^\kappa_{({\bf A},{\bf E})}( \d(t_0)\sigma^iE_{\sigma^i S_g(X)} (\gamma(t)) e^{sX}  ) \psi \right)   \Big\vert_{s=0,\kappa=0}  (x)  &=&
\nn\\&&\hspace{-4,5cm}
 S_g ( \mathds{ E} )(X(x))  \left(X(x)+\mathds{ A}(X)\right)  \psi  (x) \;,
\label{rubio}
\end{eqnarray}
where $\gamma$ is a curve generated by $X$ with $\gamma(t_0)=x$. Without the spinor and $X$ this can formally be written as $ S_g(\mathds{ E}) \nabla $, where $\nabla = d+\mathds{ A}$. 
Furthermore, if we consider a sum of operators 
$$
\hat{\cd}:=\sum_\m   \frac{d}{ds} \d(t_0) \left( \sigma^iE_{\sigma^i S_g(X_\m)} (\gamma_\m(t)) e^{sX_\m}  +    \sigma^iE_{\sigma^i S_g(-X_\m)} (\gamma_\m(t)) e^{-sX_\m}       \right)\Big\vert_{s=0} \; , \quad
$$
with  $\m\in\{1,2,3\}$ and where $X_\m=\partial_\m $ generates paths $\gamma_\m$, then we obtain
\begin{equation}
\rho^\kappa_{(\psi,{\bf E},{\bf A})}\left(   \hat{\cd}         \right) = \int_M d^3x    \bar{\psi}  \left( \nabla_\m \sigma^i\mathds{E}^\m_i  +  \sigma^i\mathds{E}^\m_i  \nabla_\m \right)  \psi + \co(\kappa)
\label{wovenhand}
\end{equation}
where we now have the classical covariant derivative $\nabla_\m= \pa_\m+\mathds{A}_\m$. The right-hand-side of (\ref{wovenhand}) is the spatial Dirac operator in three dimensions. Most significantly, notice that the background metric $g$ does not appear in (\ref{wovenhand}), which is therefore coordinate independent. Note, however, that the same statement does not hold true for the scalar product $\int \bar{\psi}\psi$ itself. See \cite{Aastrup:2009et} for a discussion of this issue\footnote{
There is another possible solution to this issue, which is interesting to consider. The question is how to deal with the determinant of the metric $e=\sqrt{g}$. A key feature of Ashtekars variables is that the triad field is densitised $E^\m_i = e e^\m_i$, where $e^\m_i$ is the triad field. This implies that the constraints also involve powers of $e$. Consider for instance the Hamilton constraint, which reads
$$
H(x) = N   \e^{ij}_{\;\;k}  E_i^\m E^\n_j F^k_{\m\n}
$$
where $N$ is the lapse field and $F_{\m\n}$ is the curvature of the Ashtekar connection. $H(x)$ comes with two factors of $e$ and must therefore be divided with one in order to be invariant, i.e. $ e^{-1} H$.

Alternatively, one may let $\psi$ in (\ref{tristese}) be a half-density. This solves the problem with the scalar product mentioned above. If one understands also the map $\rho^\kappa_{\mbox{\tiny vac}({\bf A},{\bf E})}$ in (\ref{tristese}) as an integral over half-densities, which equal the identity in the $M_2(\C)$ factor, then one can in this way also solve the issue with both the Hamilton and diffeomorphism constraints concerning the correct factors of $e$, so that the necessary factor of $e$ comes from the half-densities. This approach, however, requires a reinterpretation of the basic variables in the theory. If one is to use half-densities then one must abandon the densitised  triad field $E^\m_i $ and work directly with the triad field itself. For instance
$$
H(x) = N   \e^{ij}_{\;\;k}  e_i^\m e^\n_j F^k_{\m\n}\;.
$$
If we were to adopt this strategy it would therefore imply that we would no longer be working with Ashtekar variables and, in turn, that we would probably be dealing with a construction, which corresponds to a non-canonical quantization scheme. 
}.

Furthermore, as demonstrated in \cite{Aastrup:2010ds} we get the principal part of the Dirac Hamiltonian via a transformation $\hat{\cd}\rightarrow M(x)\hat{\cd}$ where $M(x)$ is a field, which takes values in two-by-two self-adjoint matrices and which we write as $M(x)= N(x)\mathbbm{1}+ N_i(x) \sigma^i  $, where $N,N_\m$ are the lapse and shift fields:
$$
\rightarrow \int_M d^3x    \bar{\psi} \left(  i N {\cal D}  + N^\m  \pa_\m  \right) \psi    + \mbox{\it zero-order terms}+ \co(\kappa)\;,
$$
where $N^\m= N^a E_a^\m$ and $\cd=e\sigma^i e^\m_i \nabla_\m$.

Note also that this metric independency appears to also hold for  $\kappa \not= 0$. One must, however, be careful when defining $\hat{\cd}$ because, as already mentioned, we do not have strong continuity on $\mathbf{dQH D}^* (M)  $  and therefore can, strictly speaking, not define operators using limits. This technical issue aside, lets consider the spectrum of the operator $\hat{\cd}$. Because it is independent on the background metric $g$ its resolvent appears to be compact up to an action of the diffeomorphism group.          

Finally note that the operator $\hat{\cd}$ does strictly speaking not exist in the GNS construction around the state $\rho^\kappa_{(\psi,{\bf A},{\bf E})}$ as it has been constructed so far. It is, however, straight forward to generalise the GNS construction to encompass also this operator.  \\

In \cite{Aastrup:2015gba} we defined a gravitational Hamilton operator using lattice approximations. The analysis there can also be carried over into the present lattice-independent formulation, where one will see that the gravitational Hamiltonian will i) be independent of the background metric $g$ and ii) have the correct semi-classical limit. We shall not work out the details here but simply refer the reader to  \cite{Aastrup:2015gba}, where we also discuss various critical issues concerning this operator.

\section{Discussion}

When Ashtekar in 1986 discovered a new pair of canonical variables \cite{Ashtekar:1986yd,Ashtekar:1987gu}, which now go by his name, he opened a door to a possibility of quantising general relativity using extended objects such as holonomy and flux variables. When walking through this door one is immediately confronted with the following question: {\it what operator algebra naturally incorporates Ashtekar variables both in terms of operators and in terms of a semi-classical limit?} The theory one might find depends crucially on how this question is answered. Once an algebra is chosen the theory -- if the choice is fertile! -- should follow essentially canonically via the GNS construction and some kind of canonical dynamical principle (for instance Tomita-Takesaki theory).

In a recent series of papers \cite{Aastrup:2015gba,Aastrup:2012vq,AGnew,Aastrup:2014ppa} we propose such an algebra. The $\mathbf{QH D} (M)  $ algebra, which is generated by local holonomy-diffeomorphisms and by canonical translation operators on the underlying configuration space of Ashtekar connections and which therefore essentially encodes how matter degrees of freedom are moved around on a three-dimensional manifold, incorporates Ashtekar variables both on the level of operators and -- via a natural class of states -- on the level of semi-classical approximations.

The theory which the $\mathbf{QH D} (M)  $ algebra gives rise to -- we have named it quantum holonomy theory - is, however, much more than merely an attempt to quantize general relativity. We have shown -- in these pages and in \cite{Aastrup:2015gba,Aastrup:2012vq,AGnew} -- that the theory comes with several characteristics of a unified theory of quantum gravity.  \\

The lattice-independent formulation of quantum holonomy theory reproduces most of the results found in \cite{Aastrup:2015gba} using lattice approximations. We therefore invite the reader to read also the discussion in \cite{Aastrup:2015gba} for relevant commentary. In the following we discuss a few topics, which we find particularly relevant.

At the kinematical level, we find in addition to states localised around a classical point in the configuration space of Ashtekar connections also states, which are equally distributed hereon. These two types of states are radically different: the former type appears to have vanishing overlap function whereas the latter comes with an overlap function that equals one. One might speculate if these two types of states represents a scenario where there are on the one hand high-temperature states without any notion of classical geometry and on the other hand low-temperature states, which are 'frozen' around a single classical geometry --  the two then being related by a kind of phase transition.

As a variant of the second type of states -- i.e. those without any notion of a classical geometry -- we find the possibility of a state, which is localised on a finite system of graphs. It is however unclear whether this type of states gives rise to representations of the $\mathbf{H D} (M)  $ algebra. The reason is, that seen from the Riemannian metric a graph has measure zero. Even in the case of the semiclassical states described here it is not completely settled if this gives a representation of the $\mathbf{H D} (M)  $ algebra, or if one has to construct the algebra with the counting measure, because the underlying spectrum is related to a space of "connections", which are given by the classical connection plus a singular perturbation over a path, see \cite{Aastrup:2015gba} for details. 

At the dynamical level we have demonstrated that operators, which correspond to physical quantities such as the Dirac and gravitational Hamiltonians, do not depend on the background metric used to define the $\mathbf{dQH D}^* (M)  $ algebra. More analysis is needed, however, in order to find a solid definition of the Dirac and gravitational Hamilton operators. Here a key issue to pay close attention to is the algebra of constraint operators, which must have the correct off-shell structure.

Another interesting issue is a possible connection to the framework of non-commutative geometry used to formulate the standard model of particle physics coupled to general relativity \cite{Connes:1996gi,Chamseddine:2007hz}. We have previously shown that the semi-classical limit of the $\mathbf{H D} (M)  $ algebra produces an almost-commutative algebra \cite{Aastrup:2015gba}. Furthermore, we have an operator, which produces the spatial Dirac operator in a semi-classical limit of the gravitational fields. The interesting question is whether the interaction between these two could reproduce the structure of an almost-commutative spectral triple in a semi-classical limit. If this proves to be the case we would have a very exciting connection to particle physics.

\end{document}